# THE USE OF KF STYLE FLANGES IN LOW PARTICULATE APPLICATIONS*

K. R. Kendziora[†], J. Angelo, C. Baffes, D. Franck, R. Kellett, Fermi National Accelerator Laboratory, Batavia, U.S.A.

*Abstract*

As SCRF particle accelerator technology advances the need for "low particulate" and "particle free" vacuum systems becomes greater and greater. In the course of the operation of these systems, there comes a time when various instruments have to be temporarily attached for diagnostic purposes: RGAs, leak detectors, and additional pumps. In an effort to make the additions of these instruments easier and more time effective, we propose to use KF style flanges for these types of temporary diagnostic connections. This document will describe the tests used to compare the particles generated using the assembly of the, widely accepted for "particle free" use, conflat flange to the proposed KF style flange, and demonstrate that KF flanges produce comparable or even less particles.

## INTRODUCTION

It is well established that particulate contamination can be damaging to SCRF cavities [1]. Conventionally, all-metal sealing systems with specific assembly practices are used for permanent vacuum assembly on such systems (e.g. [2]). When an accelerator with "particle free" components operates, vacuum diagnostics need to be performed. Diagnostics are used to discover the cause of strange beam behaviour, why the vacuum pressure is high, or even if the vacuum gauges are still working properly. When diagnostics need to be performed, often equipment such as RGAs, vacuum pumps, and additional gauges need to be added and then removed from the system once the diagnostics have been completed.

The preferred flange type for particle free applications at Fermilab has been a conflat flange with 316 stainless steel studs and either silicon bronze or titanium nuts. This combination provides a reliable, non-permeable seal. The drawbacks to using a conflat are the time it takes to prep the hardware, the cost of the hardware, the time required to make up the flange and more particulates are released while making up the flange. For temporary diagnostic equipment that has to be installed and removed in a relatively short time span, it would be more cost effective and time efficient to use a KF style flange. A KF style flange not only would cut down on time required to process the hardware but also in assembling the flange. This time savings becomes most critical in a "shutdown" situation where the accelerator has to be turned off for repairs, and there is a limited amount of time to perform the needed tasks. By not having to prepare 12 nuts and 6 studs (in the case of a 2.75" conflat), but only a seal and clamp, time is saved in cleaning and prep-work. Making up a conflat flange can take approximately one minute per fastener, a KF connection can be assembled in a matter of seconds.

A KF style flange can be assembled with the same, or even less particles generated in the vacuum space, as a traditional conflat flange.

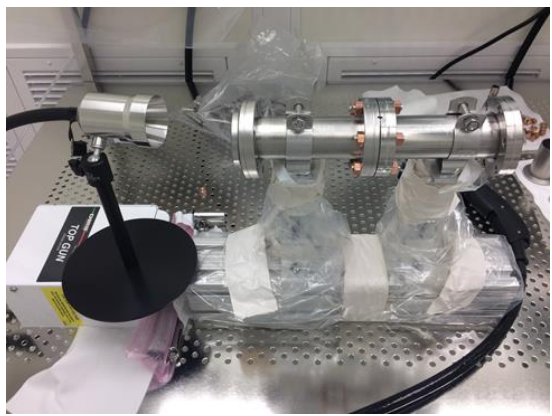

Figure 1: Test Fixture.

## TEST ENVIRONMENT AND SETUP

A 2.75" conflat and a KF-40 flange were used since they are similar in size and are fairly common interfaces for vacuum diagnostic equipment. All the hardware was cleaned ultrasonically in a 1% solution of deionized (DI) water and Micro 90 for stainless steel hardware and a 1% solution of DI water and Citranox for the Si bronze hardware. The hardware was rinsed off with clean DI water. The tests were performed in a class 10 cleanroom. The KF and conflat flanges where blown off with ionized boil-off nitrogen until 0 counts were achieved on a Climet 450t particle counter. The hardware was also blown off to the same 0 count level of cleanliness. The conflat flanges were assembled on a test stand to mitigate the introduction of particles generated from the flanges rolling around on the cleanroom bench. (see Fig. 1)

To help ensure repeatable results, a torque wrench was used to tighten the nuts on the conflat flange and the clamp on the KF flange. The conflat flange nuts were tightened to 16 Nm per the manufacturer's specification. The KF-40 flanges were tightened to 3.4 Nm to simulate finger tight since these clamps are typically tightened by hand.

---

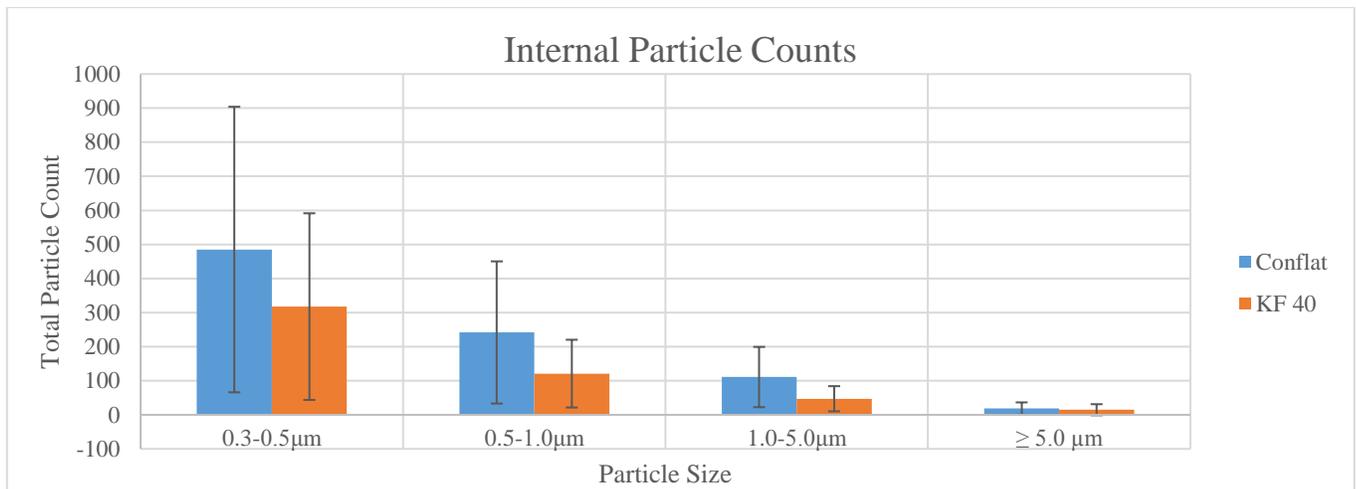

Figure 2: Internal Particle Counts. Bar height represents mean particle count across all samples. Error bar (Black bracket bars) +/- 1 standard deviation.

## PROCEDURE

Experimental and data analysis techniques largely follow the methods developed in [3]. Once the conflat hardware was blown down to a 0 particle count the studs were threaded into a nut and then inserted into the flange. After the first two studs were put into place, one hole apart, the second nut was installed to leave a gap for the copper gasket. The flange was rotated so that the two studs were in the 5 and 7 o'clock positions. The gasket was gently dropped into place. Once the gasket was in place a third stud was inserted into the top hole (12 o'clock position) and all the studs were hand tightened to hold the gasket in place. Using the torque wrench the studs were tightened to 16 Nm in a star pattern. During the stud tightening procedure, the particle counter horn was placed under the flange to trap the particles generated during the flange assembly. Once the studs were all tightened the particle counter horn was moved to be aligned with the tube and ionize boil-off nitrogen was blown through the assembly and into the particle counter until 0 counts were reached. This step was done to measure the number of particles that would be introduced into the vacuum system itself. These particles are the ones we are most concerned about because they can affect the performance of SRF cavities. Blowing out the vacuum space, under normal field conditions, is not usually an option. This procedure was followed for the remaining 19 conflat flanges. For each flange time stamps were recorded to distinguish between the different stages of the test to aid in data analysis.

### *KF-40*

The KF-40 nipples replaced the conflat spools in the test fixture and they were blown off along with the clamp, Viton o-ring and centering ring. (see Fig. 3) Once 0 counts were reached, the particle counter horn was moved to be directly under the flanges then the o-ring and center ring were carefully placed in one side of the flange. Next, the other half of the flange is gently put into place and the clamp is wrapped around the flange and tightened to 3.4 Nm. Once the clamp was tight, the particle counter horn was placed at one end of the tube and nitrogen was blown through it until 0 counts were reached. This procedure was followed for the remaining 19 flanges.

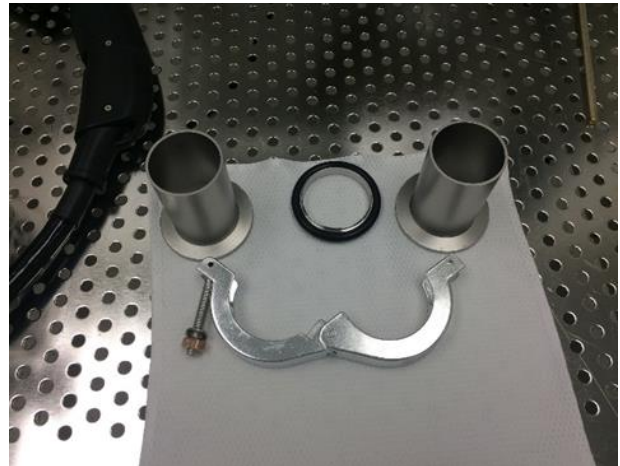

Figure 3: KF-40 nipples and hardware.

## DATA COLLECTED

After the data for 20 flanges of each type was collected some interesting observations could be made. The KF flanges produced lower mean counts on average in both the external and internal measurements. (see Figs. 2 and 4) Standard deviations are large, suggesting significant influence of outliers. The influence of outliers, particularly high-particle-count events can be seen in the histogram of sample results (see Fig. 5).

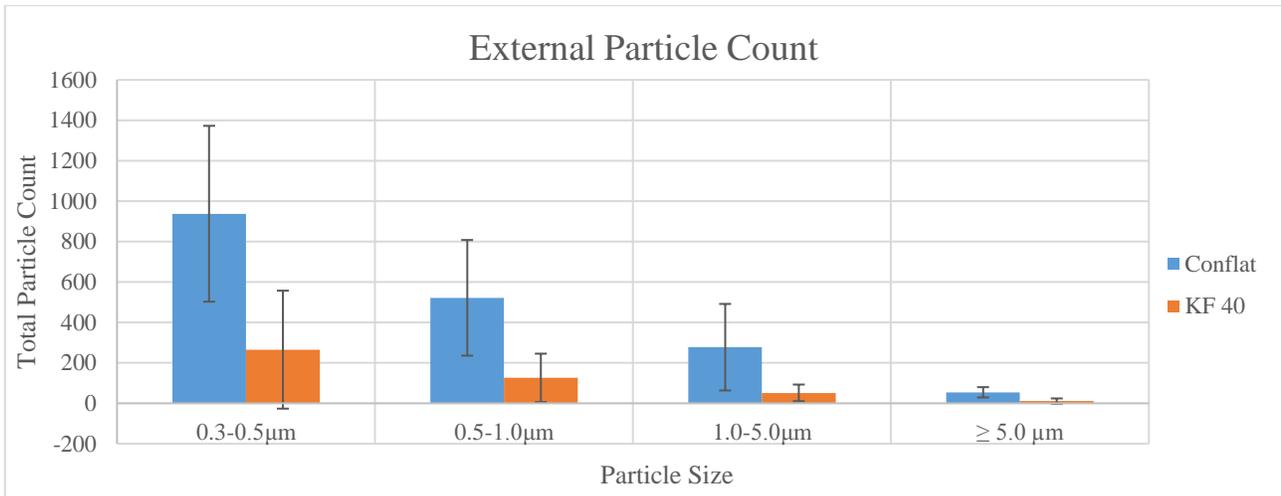

Figure 4: External particle count. Bar height represents mean particle count across all samples. Error bar +/- 1 standard deviation.

## CONCLUSION

As we can see from the data collected during this experiment a KF-40 style flange produces less overall particles and introduces less particles into the vacuum space itself when compared with the widely accepted conflat style flange. While the KF flange is a "cleaner" flange, it will not work for permanent installations due to the possibility of degradation in radiation and the fact that the Viton o-ring will permeate to some degree and limit the ultimate pressure that is achievable in a system that employs KF connections. These are the primary reasons for only using KF connections for temporary or diagnostic applications. But by using KF flanges for diagnostic connection, vacuum system down time and material cost can be minimized

## REFERENCES

[1] H. Padamsee, "RF Superconductivity: Science, Technology and Applications," Germany: Wiley-VCH, 2009

[2] S. Berry et al., "Clean Room Integration of the European XFEL Cavity Strings," in Proc. IPAC '14, Dresden, Germany, June 2014, paper WEPRI001

[3] J. Angelo, "Particle Management Strategies for Ultra-High Vacuum Assemblies," thesis, Dept. of Mechanical Engineering, Kettering University, Flint MI, USA 2016

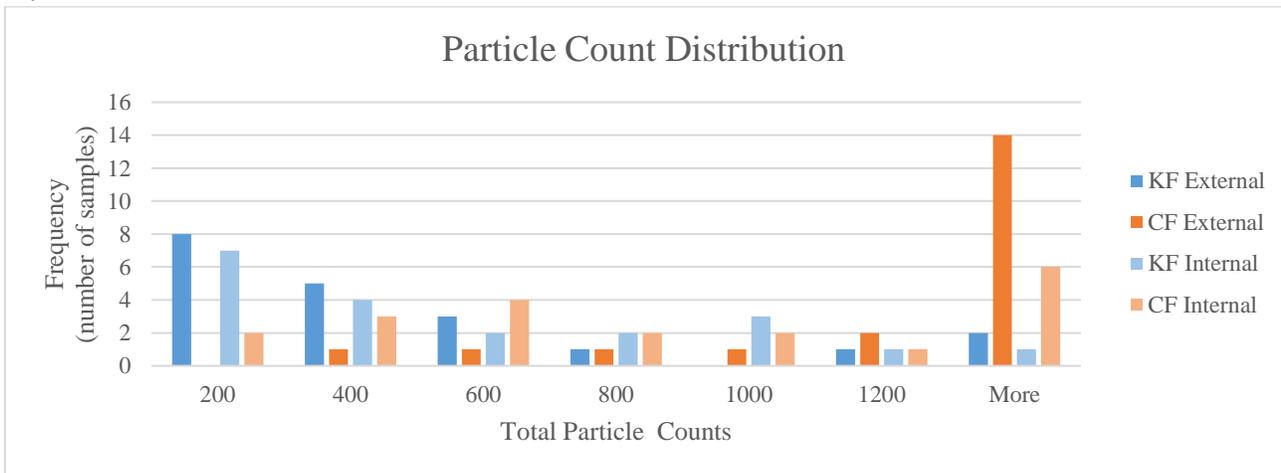

Figure 5: Particle count distribution for 20 samples, summed across all particle sizes.